\documentclass{raa}
\usepackage[a4paper, left=1in, right=1in, top=1in, bottom=1in]{geometry}

\usepackage{graphicx, times}
\usepackage{natbib}
\usepackage{ulem}

\usepackage{lipsum}
\usepackage{amssymb,amsmath}
\bibpunct{(}{)}{;}{a}{}{,}

\usepackage{orcidlink}
\usepackage{xcolor}
\usepackage{hyperref}
\hypersetup{
    colorlinks=true,
    linkcolor=blue,
    citecolor=blue,
    filecolor=magenta,
    urlcolor=cyan
}

\begin{document}

\title{Kilonova Emission from Neutron Star Mergers with Different
Equations of State}

\volnopage{Vol.0 (20xx) No.0, 000--000}

\setcounter{page}{1}

\author{Wu-Zimo Qiumu \inst{1}
\and
{Meng-Hua Chen \orcidlink{0000-0001-8406-8683}} \inst{2}
\and
{Qiu-Hong Chen \orcidlink{0009-0006-8625-5283}} \inst{1}
\and 
{En-Wei Liang \orcidlink{0000-0002-7044-733X}} \inst{1}
}

\institute{Guangxi Key Laboratory for Relativistic Astrophysics, School of Physical Science and Technology, Guangxi University, Nanning 530004, China\\
\and
Kavli Institute for Astronomy and Astrophysics, Peking University, Beijing 100871, China\\
\email{physcmh@pku.edu.cn (MHC); lew@gxu.edu.cn (EWL)}
\vs\no
{\small Received XXXX; accepted XXXX}}

\abstract{Kilonova is an optical-infrared transient powered by the radioactive decay of heavy nuclei from binary neutron star mergers. Its observational characteristics depend on the mass and the nuclide composition of meger ejecta, which are sensitive to the equation of state (EoS) of neutron star. We use astrophysical conditions derived from different EoSs as nucleosynthesis inputs to explore the impact of various EoS on the $r$-process nucleosynthesis and the kilonova emission. Our results show that both the abundance patterns of merger ejecta and kilonova light curves are strongly dependent on the neutron star EoSs. Given the mass of two neutron stars, the merger with a softer EoS tends to generate a larger amount of ejected material, and may lead to a brighter kilonova peak luminosity. The relationship between the neutron star EoS and the peak luminosity provides a probe for constraining the properties of EoS in multi-messenger observations of neutron star mergers.
}

\authorrunning{W.-Z.m. Qiumu, et al.}
\titlerunning{Kilonova with Different Equations of State}

\maketitle

\section{Introduction}          
\label{sec:intro}

It has long been considered that mergers of neutron stars or neutron star black hole systems are promising astrophysical sites for the production of heavy elements beyond iron through the rapid neutron capture process ($r$-process, \citealp{1957RvMP...29..547B,1974ApJ...192L.145L}). The radioactive decay of heavy $r$-process nuclei from merger ejecta powers an optical-infrared transient known as a kilonova \citep{1998ApJ...507L..59L,2010MNRAS.406.2650M,2012MNRAS.426.1940K,2013ApJ...775...18B,2013ApJ...774...25K,2016ApJ...829..110B,2019LRR....23....1M}. In 2017, the LIGO/Virgo collaboration discovered the first gravitational wave signal generated by binary neutron star merger (GW170817), with individual masses ranging from $1.17$ to $1.60M_{\odot}$ and a total mass of $2.74^{+0.04}_{-0.01}M_{\odot}$ \citep{2017PhRvL.119p1101A}. However, without the spin restriction, the masses of neutron stars would range between $0.86$ and $2.26M_{\odot}$ \citep{2017PhRvL.119p1101A}. Subsequently, this gravitational wave event was found to be followed by a gamma-ray burst (GRB 170817A, \citealp{2017ApJ...848L..14G}) and a kilonova (AT2017gfo, \citealp{2017ApJ...848L..12A,2017Natur.551...64A,2017Sci...358.1556C,2017ApJ...848L..17C,2017Sci...358.1570D,2017Sci...358.1565E,2017Natur.551...80K,2017Sci...358.1559K,2017Natur.551...67P,2017Sci...358.1574S,2017Natur.551...75S}). The features of the kilonova AT2017gfo are in good agreement with predictions of $r$-process kilonova models \citep{2017ApJ...848L..17C,2017Sci...358.1570D,2017Sci...358.1565E,2017Sci...358.1559K,2017ApJ...848L..18N,2017Natur.551...67P,2017Natur.551...75S,2017ApJ...848L..27T}, suggesting the synthesis of $\sim0.05M_{\odot}$ of heavy $r$-process nuclei in the ejected material. \cite{2019Natur.574..497W} analyzed the absorption features in the observed kilonova spectra and identified strontium, a heavy element with atomic number $Z=38$, in the merger ejecta. These observational evidences indicate that the mergers of two neutron stars are the primary sources of heavy $r$-process nuclei \citep{2017Natur.551...80K,2018IJMPD..2742005H,2024MNRAS.529.1154C}.

Observations of kilonova transients provide unique insight into the mass ejection from mergers and the nuclear composition of merger ejecta \citep{2017Natur.551...80K,2019Natur.574..497W,2019PhRvL.122f2701W,2021ApJ...913...26D,2022ApJ...939....8D,2023MNRAS.526L.155H,2024MNRAS.527.5540C,2024Natur.626..737L}. However, estimating the mass of the ejected material involves many systematic uncertainties, including astrophysical conditions (see \citealp{2019ARNPS..69...41S,2020ARNPS..70...95R} for reviews) and nuclear physics inputs (see \citealp{2016PrPNP..86...86M,2021RvMP...93a5002C} for reviews). In nuclear physics, properties of $r$-process nuclei, such as nuclear masses, $\beta$-decay half-lives, neutron capture rates, and fission distributions, remain unmeasured. Consequently, it remains a challenge to accurately describe the nuclear heating rate generated by the radioactive decay of $r$-process nuclei. \cite{2021ApJ...906...94Z} proposed that uncertainties in nuclear physics often lead to at least one order of magnitude variation in the inferred mass of ejected material from kilonova light curves. In astrophysics, simulations of neutron star mergers require consideration of extremely strong gravitational and magnetic fields, as well as densities and temperatures that exceed those observed in many other astrophysical phenomena. \cite{2018ApJ...869..130R} revealed that the amount and properties of the ejected material from neutron star mergers are highly sensitive to both binary parameters and the neutron star equation of state (EoS). On the one hand, the fate of the merger remnant is significantly dependent on the EoS, which determines the allowed maximum mass for neutron stars \citep{2016ARA&A..54..401O}. The central merger remnant with masses exceeding the maximum mass can collapse into black holes. Conversely, if the mass of post-merger remnant is less than the allowed maximum mass, it may be from either a massive neutron star or a stable neutron star. This has significant implications, as the central neutron star remnant can shed mass into an accretion disk, eject material via disk winds, and serve as a strong source of neutrinos, which can alter the electron fraction. On the other hand, the mass of ejected material in the presence of a neutron star remnant is also highly dependent on the EoS \citep{2013ApJ...773...78B,2013PhRvD..87b4001H,2016PhRvD..93l4046S,2018ApJ...869..130R,2024AnP...53600306R}. For example, a stiff EoS results in neutron stars of a given mass having larger radii compared to those with a soft EoS. This leads to more pronounced tidal effects, causing earlier mergers at higher orbital separations and lower orbital velocities. Additionally, stiff EoSs have higher sound speeds, making it more difficult to shock neutron star matter. Due to the less efficient shock heating and less violent post-merger oscillations, the mass ejection from neutron star mergers is significantly affected. Therefore, it is worthwhile to consider the impact of neutron star EoS on kilonova emission, as it may provide a probe for the EoS through kilonova light curves.

The relation between kilonova luminosity and neutron star EoS was investigated by \cite{2023MNRAS.522..912Z}. Without using $r$-process nucleosynthesis simulations, they adopted a semi-analytical model to calculate kilonova light curves under different EoSs and found that peak luminosities are sensitive to neutron star EoS. However, their research neglected the impact of EoS on $r$-process nucleosynthesis, including the detailed evolution process of heavy elements, nuclear heating rates, and radiative transfer processes. In this paper, we focus on exploring the impact of EoS on the nuclear composition of merger ejecta and the resulting kilonova light curve through detailed $r$-process simulations for binary neutron star mergers. By using numerical relativistic simulation results as astrophysical input for the $r$-process network, as well as detailed modeling for kilonova radiation produced by the radioactive decay of heavy $r$-process nuclei, we aim to further investigate the relationship between kilonova luminosity and EoS.

This paper is organized as follows. Section~\ref{sec:methods} details the $r$-process nucleosynthesis and the procedure to calculate the kilonova emission. Section~\ref{sec:results} presents the kilonova emission results obtained using various neutron star EoSs. Conclusions and discussions are provided in Section~\ref{sec:summary}.

\section{Methods}
\label{sec:methods}
\subsection{r-Process Nucleosynthesis}

To obtain the detailed composition of heavy $r$-process nuclei, we use the improved version of the nuclear reaction network SkyNet \citep{2015ApJ...815...82L,2017ApJS..233...18L} to perform $r$-process nucleosynthesis simulations. The nuclear physics data and the nuclear reaction rates are the same as in our previous work \citep{2023MNRAS.520.2806C,2024ApJ...971..143C,2025A&A...693A...1C}. For the radioactive decay energy data of heavy $r$-process nuclei, we use the recent database from the Evaluated Nuclear Data File library (ENDF/B-VIII.0, \citealp{Brown2018}). The astrophysical inputs for different EoSs are taken from the numerical relativity simulations provided by \cite{2018ApJ...869..130R}.

Following \cite{2024MNRAS.527.5540C}, the total $r$-process heating rate is given by
\begin{equation}
\dot{Q}(t)= f(t) \dot{q}(t),
\end{equation}
where $f(t)$ and $\dot{q}(t)$ represent the thermalization efficiency and radioactive energy generation rate, respectively. 
The radioactive decay energy released by heavy $r$-process nuclei in the merger ejecta can be written as
\begin{equation}
\dot{q}(t)= N_{\rm A} \sum_i \lambda_i E_i Y_i(t),
\end{equation}
where $\lambda_i$ is the nuclear reaction rate of the $i$th nucleus, $E_i$ is the radioactive decay energy, $Y_i(t)$ is the elemental abundance, and $N_{\rm A}$ is Avogadro's number.
The thermalization efficiency follows the analytic formula provided by \cite{2016ApJ...829..110B}
\begin{equation}
f(t) = 0.36 \left[\exp(-0.56t_{\rm day}) + \frac{\ln(1 + 0.34t_{\rm day}^{0.74})}{0.34t_{\rm day}^{0.74}}\right],
\end{equation}
where $t_{\rm day}$ is the time in days after the merger.

For the electron fraction $Y_{\rm e}$, we adopt the analytical formula fitted by \cite{2022CQGra..39a5008N}:
\begin{equation}
 Y_{\rm e}(q, \tilde{\Lambda}) = b_0 + b_1 q + b_2 \tilde{\Lambda} + b_3 q^2 + b_4 q \tilde{\Lambda} + b_5 \tilde{\Lambda}^2,
\end{equation}
where $q$ is the mass ratio of binary neutron stars, $\tilde{\Lambda}$ is the reduced tidal deformability parameter \citep{2022CQGra..39a5008N}, and $b_0$ to $b_5$ are fitting coefficients. We use the best-fit parameters provided by \cite{2022CQGra..39a5008N}. The opacity $\kappa$ as a function of $Y_{\rm e}$ is based on the results provided by \cite{2020MNRAS.496.1369T}, which were derived from a systematic analysis of the composition of heavy elements.

\subsection{Kilonova Model}

The kilonova model is based on the work of \cite{2024MNRAS.527.5540C}. We divide the ejected material into $n$ layers and the density profile can be written as \citep{2021ApJ...919...59C,2022ApJ...932L...7C}
\begin{equation}
\rho(v_n, t)=\rho_0(t)\left(\frac{v_n}{v_0}\right)^{-3},
\end{equation}
where $v_n$ is the expansion velocity of the $n$th layer.
The thermal energy of the merger ejecta evolves according to
\begin{equation}
\frac{d E_n}{d t}=-\frac{E_n}{R_n}\frac{dR_n}{dt}-L_n+\dot{Q}(t)m_n,
\end{equation}
where $R_n$ is the radius of the $n$th layer, $m_n$ is the mass of the $n$th layer, $E_n$ is the internal energy, and $L_n$ is the radiation luminosity.
The thermal luminosity of the $n$th layer is given by
\begin{equation}
L_n=\frac{E_n}{t_{{\rm lc}, n}+t_{{\rm d}, n}},
\end{equation}
where $t_{{\rm lc},n}=v_n t / c$ is the light crossing time and $t_{{\rm d}, n}=\tau_n v_n t /c$ is the photon diffusion timescale, with $\tau_n$ being the optical depth.
The total kilonova luminosity from all layers can be written as
\begin{equation}
L_{\rm bol}=\sum_n L_n.
\end{equation}

\section{Results}
\label{sec:results}

\begin{figure}
    \centering
    \includegraphics[width=0.8\textwidth]{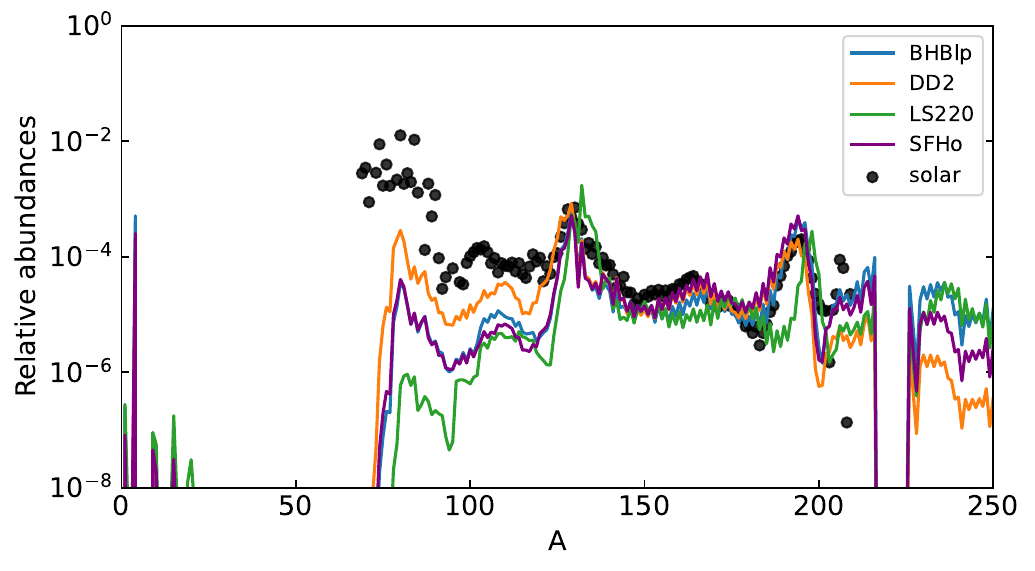}
    \caption{Abundance patterns of $r$-process nucleosynthesis calculated using different EoSs. The abundances of solar $r$-process elements taken from \cite{2007PhR...450...97A} are also shown for comparison. Astrophysical inputs for different EoSs adopted from numerical relativity simulations provided by \cite{2018ApJ...869..130R}.}
    \label{fig:1}
\end{figure}

In Figure~\ref{fig:1}, we show the resulting abundance patterns derived from $r$-process nucleosynthesis simulations for binary neutron star mergers. The astrophysical inputs for different EoSs, including BHBlp, DD2, LS220, and SFHo, are taken from numerical relativity simulations given by \cite{2018ApJ...869..130R}. It can be seen that there are differences in abundance patterns calculated using different EoSs, especially in regions with atomic mass numbers $A\ge200$ and $A\le120$. This result indicates that the EoS of neutron star plays a significant role in the $r$-process nucleosynthesis, potentially impacting the light curve of kilonova emission.

\begin{figure}
    \centering
    \includegraphics[width=0.9\textwidth]{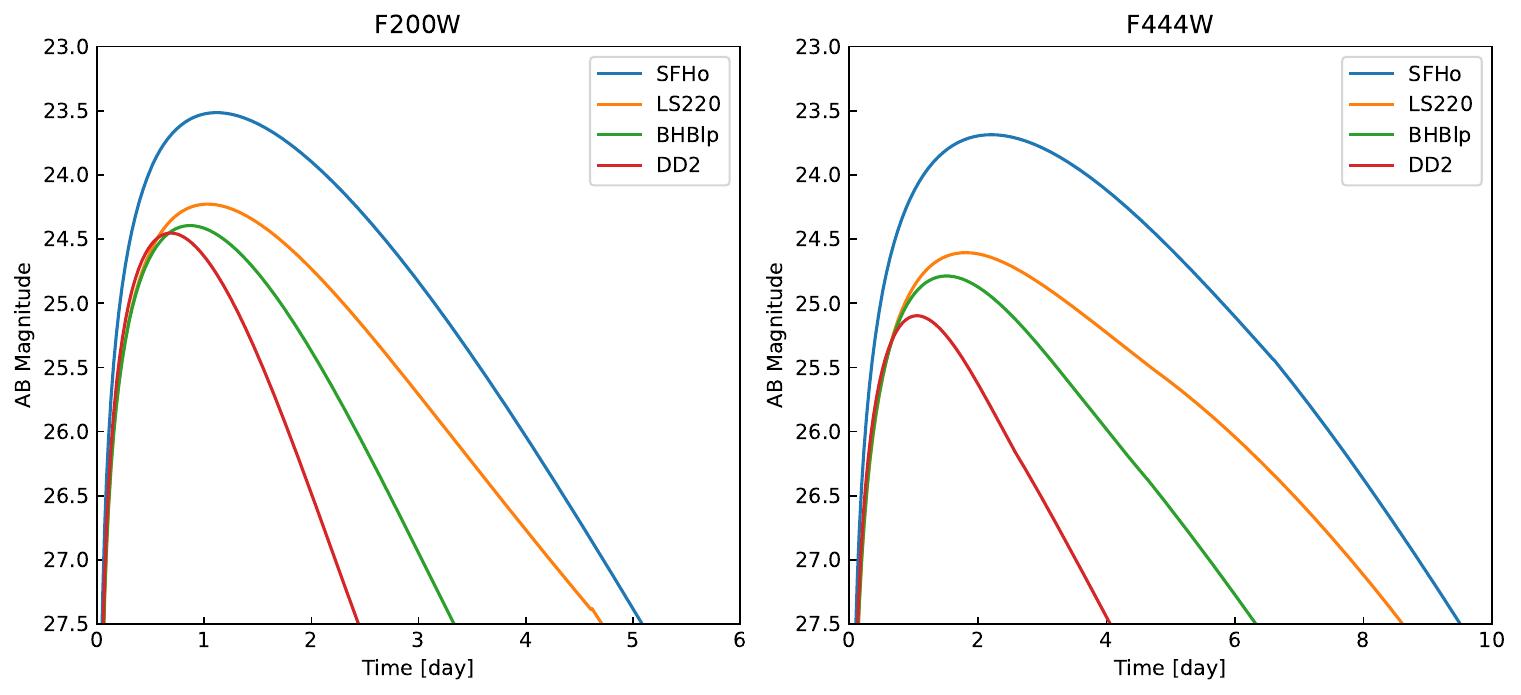}
    \caption{Kilonova light curves in the bands of F200W (left panel) and F444W (right panel) produced by the radioactive decay of $r$-process nuclei for ejecta from the merger of a symmetrical neutron star binary with the mass of $1.35+1.35M_\odot$. The blue, orange, green, and red lines represent the kilonova light curves calculated using EoS from SFHo, LS220, BHBlp, and DD2, respectively. The distance to the source is set to $200$~Mpc.
    }
    \label{fig:2}
\end{figure}

Figure~\ref{fig:2} shows the kilonova light curves produced by the radioactive decay of heavy $r$-process nuclei. Here we use the wide-band filters of the Near Infrared Camera (NIRCam) on the JWST\footnote{https://jwst-docs.stsci.edu/jwst-near-infrared-camera}. The characteristic radius ($R_{1.35}$) of a non-rotating neutron star with a mass of $1.35M_{\odot}$ calculated using the EoS from SFHo, LS220, BHBlp, and DD2 are 11.92, 12.64, 13.21 and 13.21~km, respectively. The BHBlp and DD2 predict the same radii for neutron star masses up to $1.5M_{\odot}$, but DD2 predicts a larger radius than BHBlp for masses above $1.5M_{\odot}$. Generally, an EoS with a smaller $R_{1.35}$ is regarded as `softer', whereas an EoS with a larger $R_{1.35}$ is deemed `stiffer'. Therefore, among these selected EoS, SFHo is the softest and DD2 is the stiffest. As can be seen in Figure~\ref{fig:2}, a softer EoS leads to brighter kilonova light curves and higher peak luminosity. For example, in the band of F200W, the peak luminosity calculated with soft EoS, SFHo, is higher than that calculated with stiff EoS, DD2, by a factor of $\sim2.4$ while in the band of F444W is $\sim3.7$.

\begin{figure}
    \centering
    \includegraphics[width=0.6\textwidth]{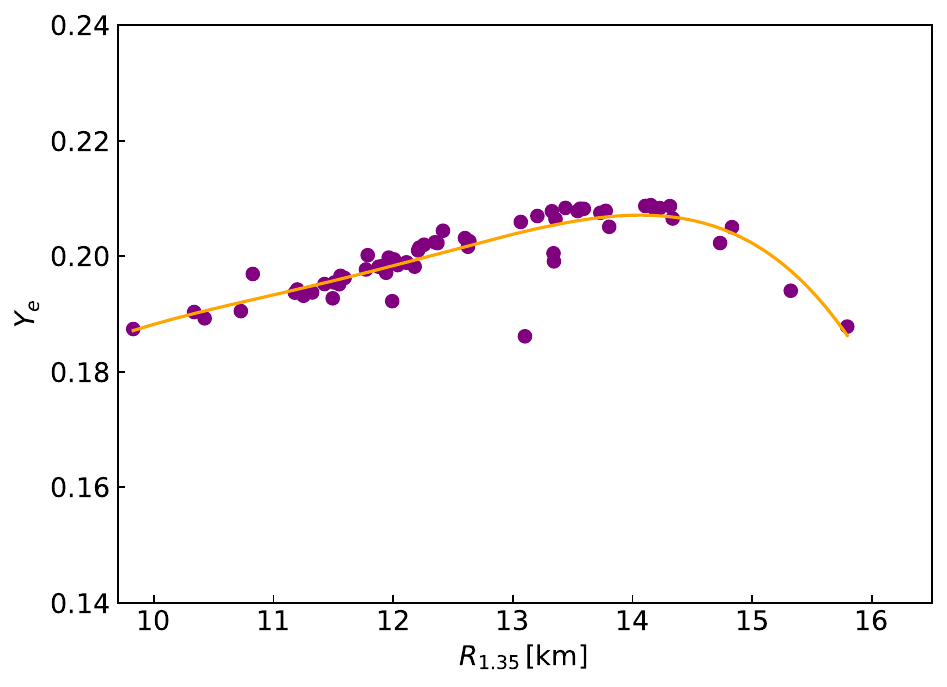}
    \caption{The electron fraction $Y_{\rm e}$ as a function of the characteristic radius $R_{1.35}$ for different EoS. Solid circles represent the results for different EoS obtained using the analytical formula from \cite{2022CQGra..39a5008N}. The solid line shows the polynomial fitting result.}
    \label{fig:3}
\end{figure}

\begin{figure}
    \centering
    \includegraphics[width=0.7\textwidth]{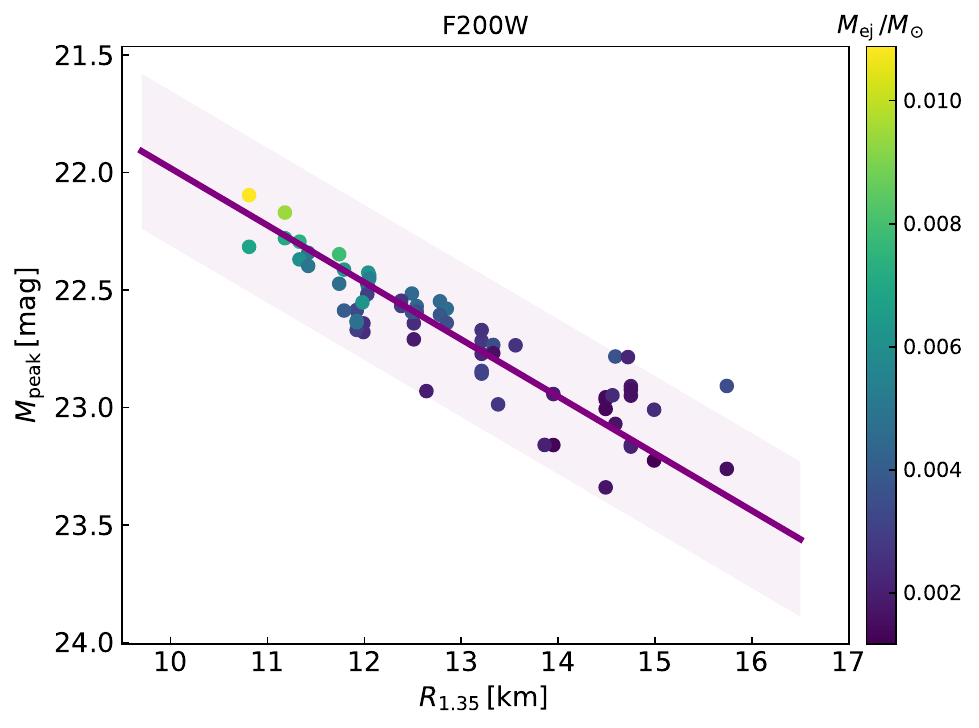} 
    \caption{The relation between peak kilonova luminosity and the characteristic radius derived from different EoSs. Solid circles indicate simulation results for a symmetric binary merger with masses of $1.35+1.35M_\odot$. The solid line represents the linear fitting result, and the colored region indicates the $\pm~1\sigma$ range.}
    \label{fig:4}
\end{figure}

To further investigate the relation between neutron star EoS and the peak luminosity, we perform $r$-process nucleosynthesis simulations using 40 distinct EoSs obtained from numerical relativistic simulations provided by \cite{2013ApJ...773...78B}. In Figure~\ref{fig:3}, we show the relationship between the electron fraction $Y_{\rm e}$ and the characteristic radius $R_{1.35}$. It is found that softer and stiff EoSs tend to produce smaller values of $Y_{\rm e}$. Figure~\ref{fig:4} shows the relationship between peak luminosity and characteristic radius $R_{1.35}$ for different EoS. Solid circles represent simulation results for a symmetric binary system with two neutron star masses of $1.35+1.35M_\odot$. It can be observed that as the characteristic radius $R_{1.35}$ decreases, the mass of the merger ejecta increases significantly, leading to a brighter peak luminosity. Our analysis indicates that the kilonova flux calculated using the softest EoS exceeds that calculated with the stiffest EoS by a factor of $\sim3.13$. This can be attributed to the fact that softer EoS have a smaller characteristic radius $R_{1.35}$, resulting in a reduction of the tidal disruption radius during neutron star mergers. This enhances the efficiency of shock heating and amplifies the kinetic energy of the oscillations. Consequently, a larger amount of merger ejecta is generated, leading to brighter kilonova emission. This indicates that, given the mass of the binary neutron star system, the observation of kilonova emission can provide information about the neutron star EoS.

\begin{figure}
    \centering
    \includegraphics[width=0.9\textwidth]{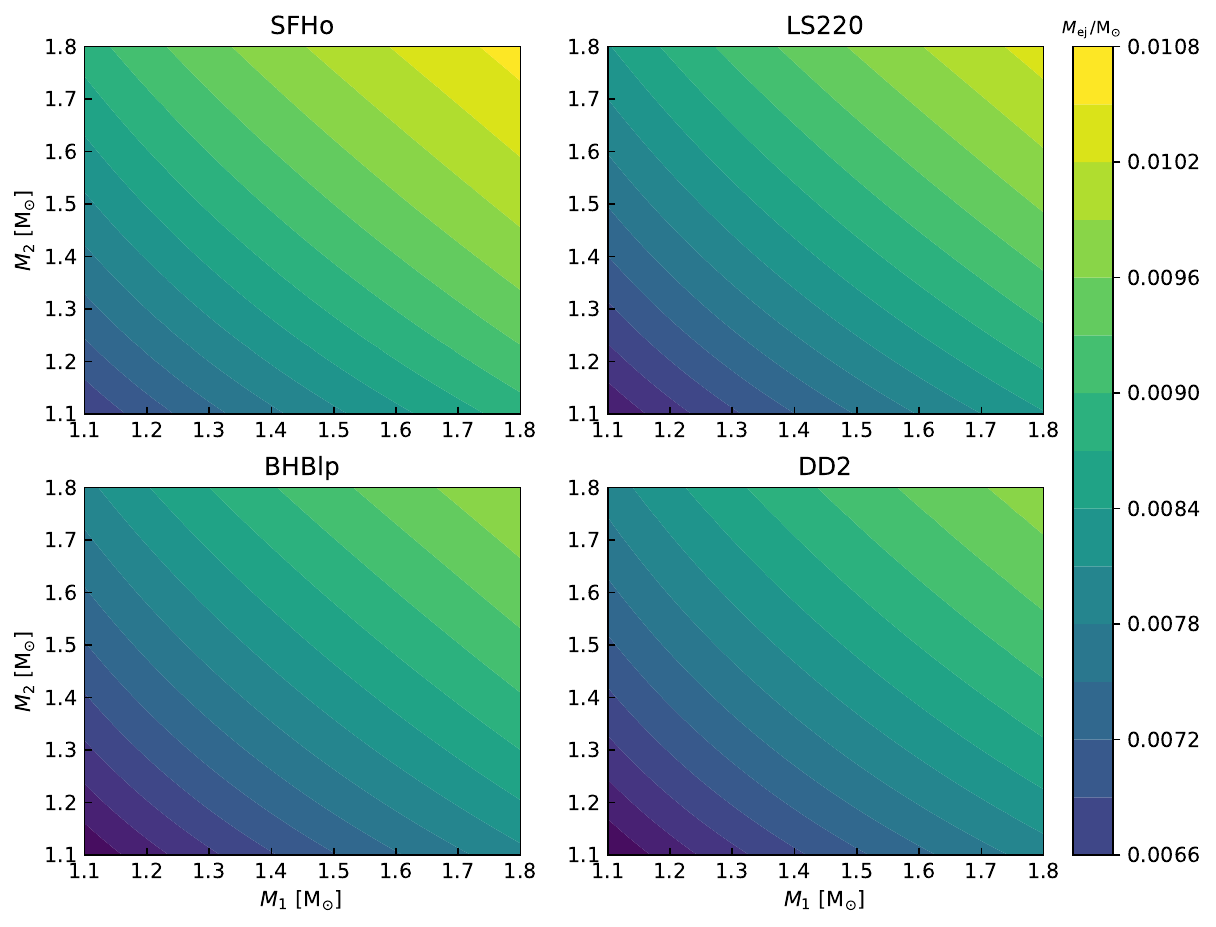}
    \caption{The mass value of merger ejecta as a function of two neutron star masses. Here we use the analytical fitting result for ejecta mass obtained from numerical relativistic simulations by \cite{2018ApJ...869..130R}.}
    \label{fig:5}
\end{figure}

\begin{figure}
    \centering
    \includegraphics[width=0.9\textwidth]{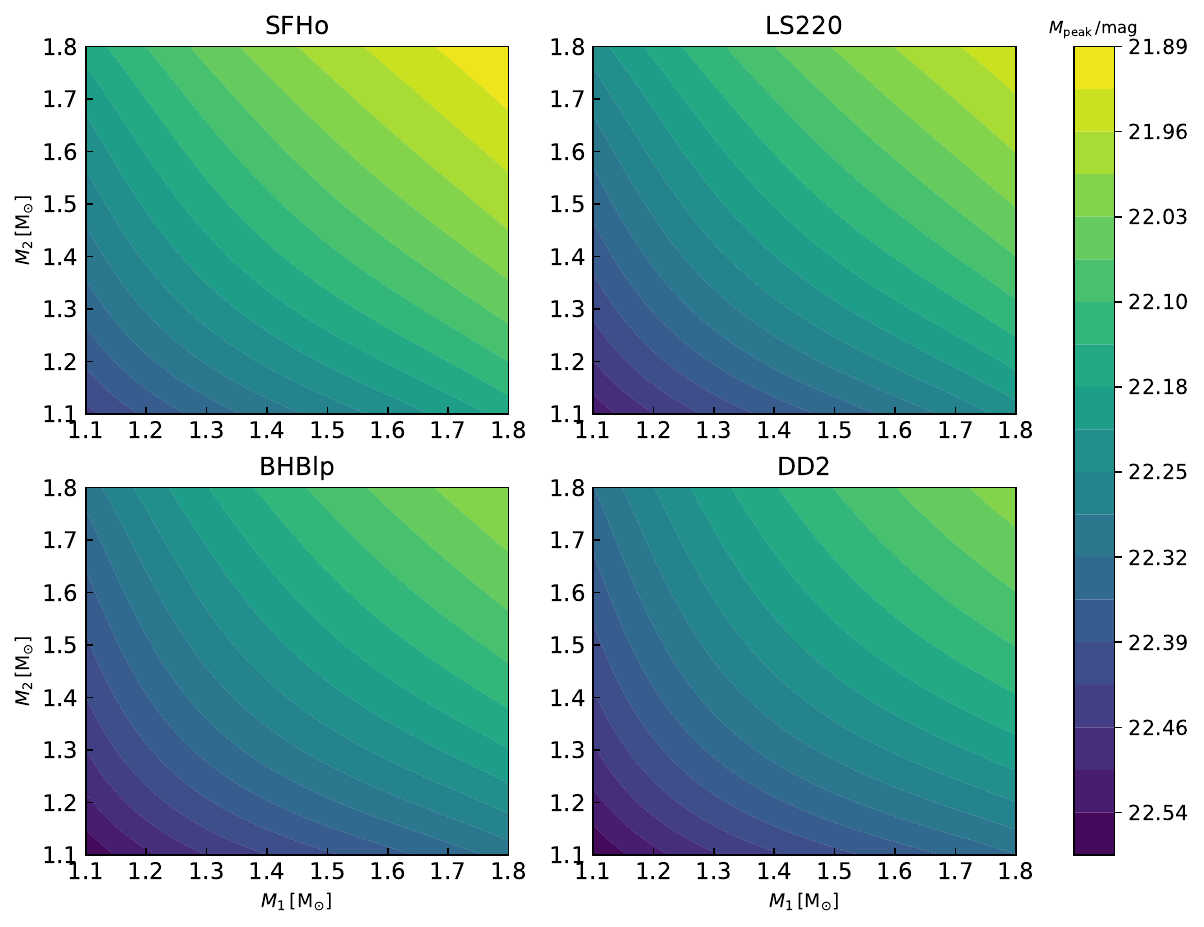}
    \caption{The same as Figure~\ref{fig:5}, but for the peak luminosity in the near-infrared band.}
    \label{fig:6}
\end{figure}

We further investigate the ejected material produced by binary neutron star mergers with different masses. We utilize the analytical fitting result for ejecta mass obtained from numerical relativistic simulations by \cite{2018ApJ...869..130R}. Figure~\ref{fig:5} shows the mass value of merger ejecta as a function of two neutron star masses obtained using four different EoSs, including SFHo, LS220, DD2, and BHBlp. It is found that in a symmetric binary system, larger neutron star masses lead to more ejected material. The neutron star EoS with softer properties such as SFHo and LS220 tend to generate a greater amount of merger ejecta compared to BHBlp and DD2. This result is consistent with the numerical relativistic simulation conducted by \citep{2013ApJ...774...25K,2013ApJ...775..113T}. Based on the ejected material from two neutron star mergers with different masses, we calculate the peak luminosity of their kilonova emission, as shown in Figure~\ref{fig:6}. It is found that the peak brightness of a kilonova increases as the binary mass increases. Notably, it can be seen that within the same binary neutron star system, a softer EoS leads to a brighter kilonova. This result suggests that kilonova emission provides a direct probe for constraining the neutron star EoS.

\section{Conclusions and Discussions}
\label{sec:summary}

In this work, we considered the neutron star EoS in binary neutron star mergers and explored its impact on the $r$-process nucleosynthesis and kilonova emission. Through detailed $r$-process nucleosynthesis simulations, we investigated the impact of the neutron star EoS on the abundance patterns of $r$-process elements. It was observed that there are differences in abundance patterns calculated using different EoSs, particularly in regions with atomic mass numbers $A\ge200$ and $A\le120$ (Figure~\ref{fig:1}). This result suggests that the neutron star EoS plays a significant role in $r$-process simulations, potentially affecting the light curve of kilonova emission.

According to the detailed composition obtained from $r$-process simulations, we calculated kilonova light curves powered by the radioactive decay of heavy nuclei (Figure~\ref{fig:2}). It is found that the peak luminosity calculated with soft EoS is higher than that calculated with stiff EoS. To further investigate the relation between neutron star EoS and the peak luminosity, we performed $r$-process nucleosynthesis simulations using 40 distinct EoSs obtained from numerical relativistic simulations. It can be observed that kilonova emission is directly related to the neutron star EoS: a softer EoS leads to brighter kilonova light curves and higher peak luminosity (Figure~\ref{fig:4}). This result is consistent with that obtained using the simple analytical formula adopted in \cite{2013ApJ...773...78B}. This can be attributed to the fact that softer EoS have a smaller characteristic radius $R_{1.35}$, resulting in a reduction of the tidal disruption radius during neutron star mergers. This enhances the efficiency of shock heating and amplifies the kinetic energy of the oscillations. Consequently, a larger amount of merger ejecta is generated, leading to brighter kilonova emission. These results are consistent with previous results \citep{2013ApJ...773...78B,2013PhRvD..87b4001H,2016PhRvD..93l4046S,2018ApJ...869..130R,2024AnP...53600306R}. We further investigate the ejected material produced by binary neutron star merger with different masses (Figures~\ref{fig:5} and \ref{fig:6}). The neutron star EoS with softer properties such as SFHo and LS220 tend to generate a greater amount of merger ejecta and power a brighter kilonova emission compared to BHBlp and DD2. This result suggests that kilonova emission provides a direct probe for constraining the neutron star EoS.

It is worth noting that our astrophysical conditions for the $r$-process nucleosynthesis calculations are obtained from numerical relativistic simulations, which typically produce merger ejecta with mass lower than those inferred from the observed kilonova light curves by about one order of magnitude \citep{2019EPJA...55..203S}. This could be due to the uncertainty in nuclear physics inputs, as the properties of heavy $r$-process nuclei are often unmeasured \citep{2021ApJ...918...44B,2021ApJ...906...94Z,2025A&A...693A...1C}. \cite{2021ApJ...906...94Z} shows that uncertainties from nuclear physics can lead to at least one order of magnitude uncertainty in kilonova luminosity. It should be noted that nuclear physics uncertainties do not affect our main conclusions, as nuclear properties are intrinsic characteristics that will influence the behavior of all kilonovae.

Multi-messenger observations of the first neutron star merger event GW170817/GRB170817A/AT2017gfo \citep{2017ApJ...848L..12A} provide a solid case for studying the neutron star EoS. The peak luminosity of kilonova AT2017gfo appears brighter than our calculated results (with a distance of $40$~Mpc), suggesting that the observed light curves support a soft EoS (i.e., smaller characteristic radius for a given mass neutron star). This is consistent with those results derived from the analysis of tidal deformability from GW170817 which suggest that the neutron star radius must be $\lesssim13$~km \citep{2018PhRvL.121p1101A,2018PhRvL.121i1102D,2018ApJ...857L..23R}. Note that we adopted a spherically symmetric model in kilonova calculations, which may affect the kilonova peak luminosity \citep{2020ApJ...897...20Z,2021ApJ...910..116K}. However, recent analysis has shown that the merger ejecta of the kilonova AT2017gfo is highly spherical, and the distribution of heavy elements is uniform \citep{2023Natur.614..436S}.

In order to effectively utilize the direct relationship between the kilonova peak luminosity and the neutron star EoS to study the properties of dense nuclear matter, the masses of two neutron stars need to be specified. The determination of neutron star masses typically depends on observations from gravitational wave observatories such as LIGO/Virgo. Given the success of the joint detection of GW170817 and AT2017gfo, there is great potential for probing the neutron star EoS based on multi-messenger analysis in the future. However, due to the degeneracy of multiple parameters, there is still some uncertainty in the neutron star masses obtained from gravitational wave signals. As the LIGO/Virgo detectors improve, the constraints on neutron star masses will be greatly enhanced. The ongoing LIGO/Virgo O4 run is expected to detect gravitational wave sources from neutron star mergers within a distance of $200$~Mpc and may detect $\sim10$ merger events. Additionally, the JWST is a powerful infrared instrument for observing kilonovae, offering a comprehensive view of kilonova emission from early to late phases \citep{2024MNRAS.527.5540C}.

\section{Acknowledgments}

We thank Li-Xin Li for the valuable discussion.
This work is supported by the National Natural Science Foundation of China (Grant Nos. 12403043, 12347172, and 12133003). M.H.C. also acknowledges support from the China Postdoctoral Science Foundation (Grant Nos. GZB20230029 and 2024M750057). This work is also supported by the Guangxi Talent Program (Highland of Innovation Talents).

\begin{center}
{\bf ORCID iDs}
\end{center}

Meng-Hua Chen: https://orcid.org/0000-0001-8406-8683

Qiu-Hong Chen: https://orcid.org/0009-0006-8625-5283

En-Wei Liang: https://orcid.org/0000-0002-7044-733X

\end{document}